\documentclass[12pt]{article}
\usepackage{psfig}

\newcommand{\be}{\begin{equation}}
\newcommand{\ee}{\end{equation}}

\newcommand{\beqa}{\begin{eqnarray}}
\newcommand{\eeqa}{\end{eqnarray}}
\newcommand{\nn}{\nonumber}

\newcommand{\eqref}[1]{(\ref{#1})}

\def\CL {{\cal L}}
\def\CM {{\cal M}}

\def\CV {{\cal V}}

\begin{document}

\setlength{\baselineskip}{7mm}
\begin{titlepage}
\begin{flushright}
{\tt NRCPS-HE-56-08} \\

December, 2008
\end{flushright}

\vspace{1cm}
\begin{center}
{\it \Large Production of charged spin-two gauge bosons \\
in \\
gluon-gluon scattering
}

\vspace{1cm}

{ \it{Spyros Konitopoulos} }
and
{ \it{George  Savvidy  } }

\vspace{0.5cm}

 {\it Institute of Nuclear Physics,} \\
{\it Demokritos National Research Center }\\
{\it Agia Paraskevi, GR-15310 Athens, Greece}

\end{center}

\vspace{1cm}

\begin{abstract}

\end{abstract}
We are considering the production of charged spin-two gauge bosons  in the  gluon-gluon
scattering and calculating polarized cross sections for each set of helicity
orientations of initial and final particles.
The angular dependence of this cross section is being compared with
the  gluon-gluon scattering cross section in QCD .

\end{titlepage}

\pagestyle{plain}

\section{\it Introduction}

An infinite tower of massive particles of high spin naturally
appears in the spectrum of different string field theories.
It is generally expected that in the tensionless limit
or, what is equivalent, at high energy and fixed angle scattering
the string spectrum becomes effectively massless
\cite{Gross:1988ue,Gross:1987kz,Witten:1988zd,
Mende:1989wt, Moore:1993ns,
Savvidy:2003fx}. In the open string theory with Chan-Paton charges these
massless states can combine  into the infinite tower
of non-Abelian tensor gauge fields \cite{Savvidy:2008ks} and one could guess that the
corresponding Lagrangian quantum field theory should be described
by some kind of extension of the Yang-Mills theory.

A possible  extension of Yang-Mills theory which includes non-Abelian tensor gauge fileds
was suggested recently in \cite{Savvidy:2005fi,Savvidy:2005zm,Savvidy:2005ki}.
The non-Abelian gauge fields are defined as rank-(s+1) tensor gauge fields
$
A^{a}_{\mu\lambda_1 ... \lambda_{s}}
$
\footnote{Tensor gauge fields
$A^{a}_{\mu\lambda_1 ... \lambda_{s}}(x),~~s=0,1,2,...$,
are totally symmetric with respect to the
indices $  \lambda_1 ... \lambda_{s}  $. {\it A priori} the tensor fields
have no symmetries with respect to the first index  $\mu$.
In particular, we have
$A^{a}_{\mu\lambda}\neq A^{a}_{\lambda\mu}$ and
$A^{a}_{\mu\lambda\rho}=A^{a}_{\mu\rho\lambda} \neq A^{a}_{\lambda\mu\rho}$.
The adjoint group index $a=1,...,N^2 -1$
in the case of $SU(N)$ gauge group. }.
The gauge invariant Lagrangian describing tensor gauge
bosons of all ranks has the form \cite{Savvidy:2005fi,Savvidy:2005zm,Savvidy:2005ki}
\be\label{Lagrangian}
{\cal L} ~= ~
{{\cal L}}_{1} + {{\cal L}}_2 +g_{3 } {{\cal L}}_3....,
\ee
where ${{\cal L}}_{1}$ is the Yang-Mills Lagrangian
and  $ {{\cal L}}_{s}~( s=2,3,..)$  are Lagrangian forms invariant
with respect to the {\it extended gauge transformations}
\cite{Savvidy:2005fi,Savvidy:2005zm,Savvidy:2005ki}.
The  Lagrangian $\CL$
defines cubic and quartic self-interactions of charged gauge quanta
carrying spin larger than one.
For the lower-rank tensor gauge fields the Lagrangian has the following form
\cite{Savvidy:2005fi,Savvidy:2005zm,Savvidy:2005ki}:
\beqa\label{totalactiontwo}
{{\cal L}}_1 =&-&{1\over 4}G^{a}_{\mu\nu}
G^{a}_{\mu\nu},\nn\\
{{\cal L}}_2  =&-&
{1\over 4}G^{a}_{\mu\nu,\lambda}G^{a}_{\mu\nu,\lambda}
-{1\over 4}G^{a}_{\mu\nu}G^{a}_{\mu\nu,\lambda\lambda}+\\
&+&{1\over 4}G^{a}_{\mu\nu,\lambda}G^{a}_{\mu\lambda,\nu}
+{1\over 4}G^{a}_{\mu\nu,\nu}G^{a}_{\mu\lambda,\lambda}
+{1\over 2}G^{a}_{\mu\nu}G^{a}_{\mu\lambda,\nu\lambda},\nn
\eeqa
where the generalized field strength tensors are:
\beqa\label{tensors}
G^{a}_{\mu\nu} &=&
\partial_{\mu} A^{a}_{\nu} - \partial_{\nu} A^{a}_{\mu} +
g f^{abc}~A^{b}_{\mu}~A^{c}_{\nu}, \nn\\
G^{a}_{\mu\nu,\lambda} &=&
\partial_{\mu} A^{a}_{\nu\lambda} - \partial_{\nu} A^{a}_{\mu\lambda} +
g f^{abc}(~A^{b}_{\mu}~A^{c}_{\nu\lambda} + A^{b}_{\mu\lambda}~A^{c}_{\nu} ~), \\
G^{a}_{\mu\nu,\lambda\rho} &=&
\partial_{\mu} A^{a}_{\nu\lambda\rho} - \partial_{\nu} A^{a}_{\mu\lambda\rho} +
g f^{abc}(~A^{b}_{\mu}~A^{c}_{\nu\lambda\rho} +
 A^{b}_{\mu\lambda}~A^{c}_{\nu\rho}+A^{b}_{\mu\rho}~A^{c}_{\nu\lambda}
 + A^{b}_{\mu\lambda\rho}~A^{c}_{\nu} ~) .\nn
\eeqa
The definition of
the Lagrangian forms $ {{\cal L}}_{s}$
for higher-rank  fields can be found in the previous
publications \cite{Savvidy:2005fi,Savvidy:2005zm,Savvidy:2005ki}.
The above expressions define interacting gauge field theory with infinite
many gauge fields. Not much is known about physical properties of such gauge
field theories and in the present paper we shall focus our attention on the lower-rank
tensor gauge field $A^{a}_{\mu\lambda}$, which
describes in this theory charged gauge bosons of spin two. We are interested
in studying  the first nontrivial interaction processes.
In particular, we shall consider
production of  charged spin-two gauge bosons by a pair of vector gauge bosons.

Our intention in this article is to calculate the leading-order differential
cross section of spin-two tensor gauge boson production by a pair of vector
gauge bosons in the process $V+V  \rightarrow T + T$ and to analyze the
angular dependence of the polarized cross sections
for each set of helicity orientations of initial and final particles.
The process is illustrated in   Fig.\ref{fig1} and
receives contribution from four Feynman diagrams shown in  Fig.\ref{fig5}-Fig.\ref{fig8}.

Below we shall present the Feynman diagrams for the given process,
the expressions for the corresponding vertices and transition amplitudes.
Then we shall calculate the polarized cross sections for each set of helicity orientations of the initial
and final particles (see formulas
(\ref{crosssectionformulaLLLL}), (\ref{crosssectionformulaLLRR}),
(\ref{crosssectionformulaLRLL}) and (\ref{crosssectionunpolarized}))  and
shall compare them with the corresponding
cross section of the vector gauge bosons $V+V \rightarrow V+V$  in Yang-Mills theory
(see formulas (\ref{sqvvvv1}), (\ref{sqvvvv2}), (\ref{sqvvvv3}) and (\ref{VVVVscattering})).
In Appendix A we are reviewing the well known result for the three-level scattering
$V+V \rightarrow V+V$ \cite{Peskin:1995ev} and in Appendix B
we shall demonstrate the gauge invariance of the transition amplitude.

\section{\it Summary of Feynman rules }

The Feynman rules for the Lagrangian (\ref{Lagrangian}) can be
derived from the functional
integral over the  gauge boson fields
$A^{a}_{\alpha},~A^{a}_{\alpha\alpha'},...$ \cite{Savvidy:2005fi,Savvidy:2005zm,Savvidy:2005ki}.
The indices of the symmetry group G are $a,b=1,...,d(G)$, where $d(G)$ is the number
of generators of the group G.
The standard vector propagator is given by the expression
\beqa\label{vectorpropagator}
D^{\alpha\beta}_{ab}(k)=-{i\over k^{2}}\eta^{\alpha\beta}\delta_{ab}.
\eeqa
The second-rank tensor gauge field $A_{\alpha\acute{\alpha}}$ with 16
components  describes in this theory three
physical transversal polarizations \cite{Savvidy:2005fi,Savvidy:2005zm,Savvidy:2005ki}.
The kinetic operator of the tensor gauge bosons
\beqa\label{basickineticterm}
H_{\alpha\acute{\alpha}\gamma\acute{\gamma}}(k)&=&
(-\eta_{\alpha\gamma}\eta_{\acute{\alpha}\acute{\gamma}}
+{1 \over 2}\eta_{\alpha\acute{\gamma}}\eta_{\acute{\alpha}\gamma}
+{1 \over 2}\eta_{\alpha\acute{\alpha}}\eta_{\gamma\acute{\gamma}})k^2
+\eta_{\alpha\gamma}k_{\acute\alpha}k_{\acute{\gamma}}
+\eta_{\acute\alpha \acute{\gamma}}k_{\alpha}k_{\gamma}\nn\\
&-&{1 \over 2}(\eta_{\alpha\acute{\gamma}}k_{\acute\alpha}k_{\gamma}
+\eta_{\acute\alpha\gamma}k_{\alpha}k_{\acute{\gamma}}
+\eta_{\alpha\acute\alpha}k_{\gamma}k_{\acute{\gamma}}
+\eta_{\gamma\acute{\gamma}}k_{\alpha}k_{\acute\alpha})
\eeqa
is a gauge invariant operator
$
k_{\alpha}H_{\alpha\acute{\alpha}\gamma\acute{\gamma}}=0,~
k_{\acute{\alpha}}H_{\alpha\acute{\alpha}\gamma\acute{\gamma}}=0.
$
It describes the propagation of massless particles with helicities two
and zero because the equation
$
H_{\alpha\acute{\alpha}\gamma\acute{\gamma}}(k) f^{\gamma\acute{\gamma}}(k)=0
$
has three independent solutions of the helicity two and zero.
The propagator $\Delta^{ab}_{\alpha\alpha'\beta\beta'}(k)$
is defined through the equation
$
H^{fix}_{\alpha\acute{\alpha}\gamma\acute{\gamma}}(k)
\Delta^{\gamma\acute{\gamma}}_{~~~\lambda\acute{\lambda}}(k) =
i \eta_{\alpha\lambda}\eta_{\acute{\alpha}\acute{\lambda}}~~
$
and has the following form:
\be\label{tensorpropagator}
\Delta^{ab}_{\alpha\alpha'\beta\beta'}(k) = -
{i  \over k^2  } ~\Pi_{\alpha\alpha'\beta\beta'}~
\delta^{ab},
\ee
where the residue can be represented as a sum of $\lambda= \pm 2$ and
$\lambda= 0$ helicity states:
\beqa
\Pi_{\alpha\alpha'\beta\beta'} =
(\eta_{\alpha\beta} \eta_{\alpha'\beta'}
+\eta_{\alpha \beta'}\eta_{\alpha' \beta}
-\eta_{\alpha \alpha'} \eta_{\beta\beta'})+
 {1\over 3}(\eta_{\alpha\beta}\eta_{\alpha'\beta'}
-\eta_{\alpha\beta'}\eta_{\alpha'\beta})  .
\eeqa
\begin{figure}
\centerline{\hbox{\psfig{figure=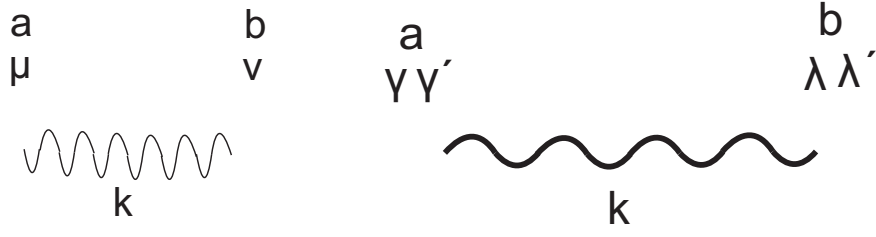,height=3cm,angle=0}}}
\caption[fig2]{The vector - $D^{ab}_{\mu\nu}(k)$ and tensor -
$\Delta^{ab}_{\gamma\gamma^{'}\lambda\lambda^{'} }(k)$ gauge boson propagators
are conventionally drawn as thin and thick wave lines.}
\label{fig2}
\end{figure}
The standard Yang-Mills  three-vector boson interaction vertex
VVV in the momentum representation has the form
\be\label{VVV}
{{\cal V}}^{abc}_{\alpha\beta\gamma}(k,p,q)= -  g f^{abc}
 F_{\alpha\beta\gamma}(k,p,q) =
-  g f^{abc} [\eta_{\alpha\beta} (p-k)_{\gamma}+ \eta_{\alpha\gamma} (k-q)_{\beta}
 + \eta_{\beta\gamma} (q-p)_{\alpha}] .
\ee
The interaction vertex of vector gauge boson V with two
tensor gauge bosons T - the VTT vertex - has the form\footnote{See formulas
(62),(65) and (66) in \cite{Savvidy:2005ki} .}
 \cite{Savvidy:2005zm,Savvidy:2005ki}
\be\label{VTT}
\CV^{abc}_{\alpha\acute{\alpha}\beta\gamma\acute{\gamma}}(k,p,q) =
- g  ~ f^{abc} F_{\alpha\acute{\alpha}\beta\gamma\acute{\gamma}} ,
\ee
where
\beqa\label{vertexoperator1}
F_{\alpha\acute{\alpha}\beta\gamma\acute{\gamma}}(k,p,q)
&=& [\eta_{\alpha\beta} (p-k)_{\gamma}+ \eta_{\alpha\gamma} (k-q)_{\beta}
 + \eta_{\beta\gamma} (q-p)_{\alpha}] \eta_{\acute{\alpha}\acute{\gamma}}- \\
&-&{1\over 2} [~(p-k)_{\gamma}(\eta_{\alpha\acute{\gamma}}
\eta_{\acute{\alpha}\beta}+
\eta_{\alpha\acute{\alpha}} \eta_{\beta\acute{\gamma}})
+ (k-q)_{\beta}(\eta_{\alpha\acute{\gamma}} \eta_{\acute{\alpha}\gamma}+
\eta_{\alpha\acute{\alpha}} \eta_{\gamma\acute{\gamma}})\nn\\
&+& (q-p)_{\alpha} (\eta_{\acute{\alpha}\gamma} \eta_{\beta\acute{\gamma}}+
\eta_{\acute{\alpha}\beta} \eta_{\gamma\acute{\gamma}})
+(p-k)_{\acute{\alpha}}\eta_{\alpha\beta} \eta_{\gamma\acute{\gamma}}+
(p-k)_{\acute{\gamma}} \eta_{\alpha\beta} \eta_{\acute{\alpha}\gamma}\nn\\
&+&(k-q)_{\acute{\alpha}} \eta_{\alpha\gamma} \eta_{\beta\acute{\gamma}}+
(k-q)_{\acute{\gamma}}\eta_{\alpha\gamma} \eta_{\acute{\alpha}\beta}
+(q-p)_{\acute{\alpha}} \eta_{\beta\gamma} \eta_{\alpha\acute{\gamma}}+
(q-p)_{\acute{\gamma}}\eta_{\alpha\acute{\alpha}} \eta_{\beta\gamma} ].\nn
\eeqa
The Lorentz indices $\alpha\acute{\alpha}$ and momentum $k$ belong to the
first tensor gauge boson, the $\gamma\acute{\gamma}$ and momentum $q$
belong to the second tensor gauge boson, and Lorentz index $\beta$  and
momentum $p$ belong to the vector gauge boson. The vertex is shown in
Fig.\ref{fig3}. Vector gauge bosons are conventionally drawn
as thin wave lines, tensor gauge bosons are thick wave lines.

\begin{figure}
\centerline{\hbox{\psfig{figure=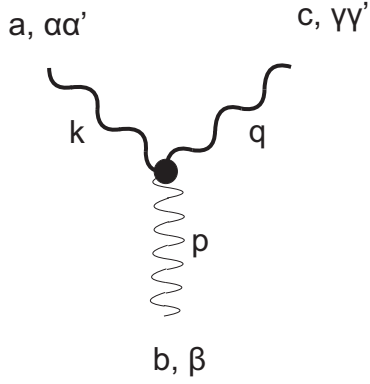,height=5cm,angle=0}}}
\caption[fig3]{The interaction vertex for vector gauge boson V and
two tensor gauge bosons T
the VTT vertex - $\CV^{abc}_{\alpha\acute{\alpha}\beta\gamma\acute{\gamma}}(k,p,q)$
in non-Abelian tensor gauge
field theory \cite{Savvidy:2005ki}.
Vector gauge bosons are conventionally drawn
as thin wave lines, tensor gauge bosons are thick wave lines.
The Lorentz indices $\alpha\acute{\alpha}$ and momentum $k$ belong to the
first tensor gauge boson, the $\gamma\acute{\gamma}$ and momentum $q$
belong to the second tensor gauge boson, and Lorentz index $\beta$  and
momentum $p$ belong to the vector gauge boson. }
\label{fig3}
\end{figure}

In the Lagrangian
(\ref{totalactiontwo}) we have the
standard  four vector boson interaction vertex VVVV
\beqa\label{VVVV}
{{\cal V}}^{abcd}_{\alpha\beta\gamma\delta}(k,p,q,r) = -i g^2 f^{lac}f^{lbd} (\eta_{\alpha \beta}
\eta_{\gamma\delta} - \eta_{\alpha \delta} \eta_{\beta\gamma})\nonumber\\
-ig^2 f^{lad}f^{lbc} (\eta_{\alpha \beta} \eta_{\gamma\delta} -
\eta_{\alpha \gamma}\eta_{\beta\delta} )\nonumber\\
-ig^2 f^{lab}f^{lcd} (\eta_{\alpha \gamma} \eta_{\beta\delta} -
\eta_{\alpha \delta}\eta_{\beta\gamma} )
\eeqa
and a new interaction of two vector and two tensor gauge bosons - the VVTT vertex,
\beqa\label{VVTT}
{{\cal V}}^{abcd}_{\alpha\beta\gamma\acute{\gamma}\delta\acute{\delta}}(k,p,q,r)=
&-&  ig^2    ~f^{lac}f^{lbd} (\eta_{\alpha \beta}
\eta_{\gamma\delta} - \eta_{\alpha \delta} \eta_{\beta\gamma})
\eta_{\acute{\gamma}\acute{\delta}}\nonumber\\
&-& ig^2   ~f^{lad}f^{lbc} (\eta_{\alpha \beta} \eta_{\gamma\delta} -
\eta_{\alpha \gamma}\eta_{\beta\delta} )\eta_{\acute{\gamma}\acute{\delta}}\nonumber\\
&-& ig^2    ~f^{lab}f^{lcd} (\eta_{\alpha \gamma} \eta_{\beta\delta} -
\eta_{\alpha \delta}\eta_{\beta\gamma} )\eta_{\acute{\gamma}\acute{\delta}} \nn\\
+ {i \over 2}g^2    ~f^{lac}f^{lbd} [&+&\eta_{\alpha \beta}
(\eta_{\gamma\acute{\delta}}\eta_{\acute{\gamma}\delta} +
\eta_{\gamma\acute{\gamma}} \eta_{\delta\acute{\delta}})\nn\\
&-&\eta_{\beta\gamma}
(\eta_{\alpha \acute{\delta}}\eta_{\acute{\gamma}\delta} +
\eta_{\alpha\acute{\gamma}} \eta_{\delta\acute{\delta}})\nn\\
&-&\eta_{\alpha\delta}
(\eta_{\beta\acute{\gamma}}\eta_{\gamma\acute{\delta}} +
\eta_{\beta\acute{\delta}} \eta_{\gamma\acute{\gamma}})\nn\\
&+&\eta_{\gamma\delta}
(\eta_{\alpha\acute{\delta}}\eta_{\beta\acute{\gamma}} +
\eta_{\alpha\acute{\gamma}} \eta_{\beta\acute{\delta}})]\nn\\
+{i \over 2}g^2    ~f^{lad}f^{lbc} [&+&\eta_{\alpha \beta}
(\eta_{\gamma\acute{\delta}}\eta_{\acute{\gamma}\delta} +
\eta_{\gamma\acute{\gamma}} \eta_{\delta\acute{\delta}})\nn\\
&-&\eta_{\alpha\gamma}
(\eta_{\beta\acute{\delta}}\eta_{\acute{\gamma}\delta} +
\eta_{\beta\acute{\gamma}} \eta_{\delta\acute{\delta}})\nn\\
&-&\eta_{\beta\delta}
(\eta_{\alpha\acute{\gamma}}\eta_{\gamma\acute{\delta}} +
\eta_{\alpha\acute{\delta}} \eta_{\gamma\acute{\gamma}})\nn\\
&+&\eta_{\gamma\delta}
(\eta_{\alpha\acute{\gamma}}\eta_{\beta\acute{\delta}} +
\eta_{\alpha\acute{\delta}} \eta_{\beta\acute{\gamma}})]\nn\\
+{i \over 2}g^2  ~ f^{lab}f^{lcd}[&+&\eta_{\alpha \gamma}
(\eta_{\beta\acute{\gamma}}\eta_{\delta\acute{\delta}} +
\eta_{\beta\acute{\delta}} \eta_{\delta\acute{\gamma}})\nn\\
&-&\eta_{\beta\gamma}
(\eta_{\alpha\acute{\gamma}}\eta_{\delta\acute{\delta}} +
\eta_{\alpha\acute{\delta}} \eta_{\delta\acute{\gamma}})\nn\\
&-&\eta_{\alpha\delta}
(\eta_{\beta\acute{\delta}}\eta_{\gamma\acute{\gamma}} +
\eta_{\beta\acute{\gamma}} \eta_{\gamma\acute{\delta}})\nn\\
&+&\eta_{\beta\delta}
(\eta_{\alpha\acute{\delta}}\eta_{\gamma\acute{\gamma}} +
\eta_{\alpha\acute{\gamma}} \eta_{\gamma\acute{\delta}})].
\eeqa
In summary, we have the Yang-Mills vertex VVV (\ref{VVV}), the new
vertex  VTT (\ref{VTT}) together with the Yang-Mills vertex VVVV (\ref{VVVV})
and the new vertex VVTT (\ref{VVTT}) (see Fig.{\ref{fig3},{\ref{fig4}).
\begin{figure}
\centerline{\hbox{\psfig{figure=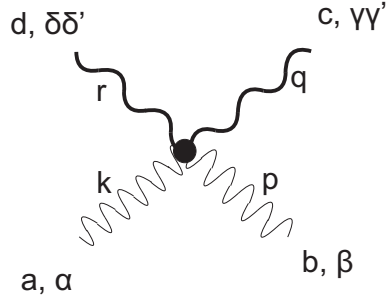,width=5cm}}}
\caption[fig4]{The quartic vertex with two vector gauge bosons and two
tensor gauge bosons - the VVTT vertex -
${{\cal V}}^{abcd}_{\alpha\beta\gamma\acute{\gamma}\delta\acute{\delta}}(k,p,q,r)$
in non-Abelian tensor gauge
field theory \cite{Savvidy:2005ki}.
Vector gauge bosons are conventionally drawn
as thin wave lines, tensor gauge bosons are thick wave lines.
The Lorentz indices $\gamma\acute{\gamma}$ and momentum $q$ belong to the
first tensor gauge boson, $\delta\acute{\delta}$ and momentum $r$ belong to the
second tensor gauge boson, the index $\alpha$  and momentum $k$
belong to the first vector gauge boson and Lorentz index $\beta$  and
momentum $p$ belong to the second vector gauge boson.
}
\label{fig4}
\end{figure}

\section{\it Cross section and Matrix Elements }
\begin{figure}
\centerline{\hbox{\psfig{figure=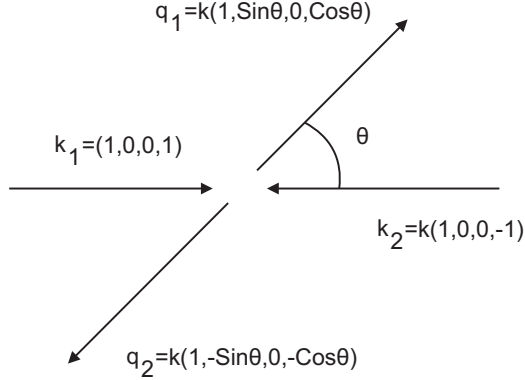,height=5cm,angle=0}}}
\caption[fig1]{The scattering $V+V  \rightarrow T+ T$, shown in the
center-of-mass frame. The $k_{1},k_2$ are momenta of the vector gauge bosons $V$ and $q_{1}, q_2$
are momenta of the tensor gauge bosons $T$.}
\label{fig1}
\end{figure}
The scattering process is illustrated in   Fig.\ref{fig1}.
Working in the center-of-mass frame, we make the following assignments:
$
k_{1}^{\mu}=E(1,0,0,1),~~ k_{2}^{\mu}=E(1,0,0,-1),
$~~and
$q_{1}^{\mu}=E(1,\sin\theta,0,\cos\theta),~~
q_{2}^{\mu}=E(1,-sin\theta,0,-\cos\theta),$
where $k_{1,2}$ are momenta of the vector bosons $V+V$ and $q_{1,2}$
momenta of the tensor gauge bosons $T+T$. All particles are massless
$k_{1}^{2} = k_{2}^{2} = q^{2}_1  = q^{2}_2  =0 $.
In the center-of-mass frame
the momenta satisfy the relations $\vec{k}_1  = -\vec{k}_2$, $\vec{q}_2  = -\vec{q}_1$.
The invariant variables of the process are:
\beqa
s =  2 (k_1 \cdot k_2),~~~
t=  -{s\over 2} (1-\cos \theta ),~~~
u= -{s\over 2} (1+\cos \theta ) \nn,
\eeqa
where $s= (2E)^2$ and $\theta$ is the scattering angle.
It is convenient to write the differential cross section in the center-of-mass frame with
tensor boson produced into the solid angle $d \Omega$ as
\be\label{crosssectionformula}
d\sigma = {1 \over 2 s} \vert M \vert^2 {1\over 32 \pi^2} d\Omega,
\ee
where the final-state density  is
$
d \Phi =   {1\over 32 \pi^2} d\Omega .
$

We shall calculate the polarized cross sections for the reaction $V+V \rightarrow T+T$, to the lowest order
in $\alpha = g^2 / 4\pi$. The lowest-order Feynman diagrams contributing to
the annihilation process of a pear of vector bosons into a pair of tensor gauge
bosons are shown in Fig.\ref{fig5}-Fig.\ref{fig8}. In order $g^2 $,
there are five diagrams.
Vector gauge bosons  $V $ are conventionally drawn
as thin wave lines and tensor gauge bosons $T$ as a thick wave lines.
\begin{figure}
\centerline{\hbox{\psfig{figure=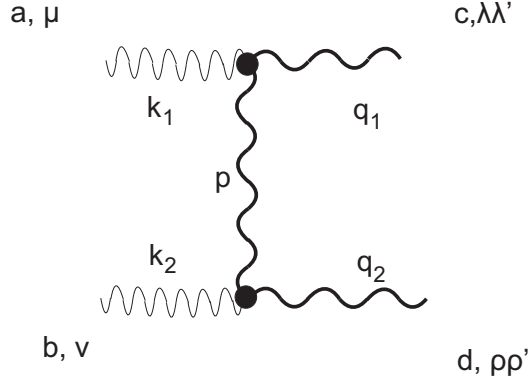,height=5cm,angle=0}}}
\caption[fig5]{The t-channel diagram corresponding to the creation
of tensor gauge bosons by vector bosons $V+V \rightarrow T+T$.}
\label{fig5}
\end{figure}

The probability amplitude of the process can be written as a sum of four terms
corresponding to each diagram. For the first diagram Fig.\ref{fig5} we shall get
\beqa\label{polarizedtransitionamplitudeI}
&i\CM_{I}^{ab,cd}=\\
=&\CV_{\lambda\lambda'\mu~\sigma\sigma'}^{cae}(-q_{1},k_{1},-p)~
\Delta^{\sigma\sigma'\tau\tau'}_{el}(p)~
\CV_{\tau\tau'\nu~\rho\rho'}^{lb d}(p,k_{2},-q_{2})~
e^{\mu}_{k_{1}}~ e^{\nu}_{k_{2}}~ e^{*\lambda\lambda'}_{q_{1}} ~e^{*\rho\rho'}_{q_{2}}\nn,
\eeqa
for the second diagram Fig.\ref{fig6} the amplitude is
\beqa\label{polarizedtransitionamplitudeII}
&i\CM_{II}^{ab,cd}=\\
=&\CV_{\rho\rho'\mu~\sigma\sigma'}^{dae}(-q_{2},k_{1},-p')~
\Delta^{\sigma\sigma'\tau\tau'}_{el}(p')~
\CV_{\tau\tau'\nu~\lambda\lambda'}^{lbc}(p',k_{2},-q_{1}) ~
e^{\mu}_{k_{1}}~e^{\nu}_{k_{2}}~e^{*\lambda\lambda'}_{q_{1}}~e^{*\rho\rho'}_{q_{2}}\nn,
\eeqa
for the third  diagram Fig.\ref{fig7} it is
\beqa\label{polarizedtransitionamplitudeIII}
&i\CM_{III}^{ab,cd}=\\
=&\CV_{\mu\nu\sigma}^{abe}(k_{1},k_{2},-p'')~
D^{\sigma\tau}_{el}(p'')~
\CV_{\lambda\lambda'\tau\rho\rho'}^{cld}(-q_{1},p'',-q_{2})~
e^{\mu}_{k_{1}}~e^{\nu}_{k_{2}}~e^{*\lambda\lambda'}_{q_{1}}~e^{*\rho\rho'}_{q_{2}} \nn
\eeqa
and finally for the forth diagram Fig.\ref{fig8} we have
\beqa\label{polarizedtransitionamplitudeIV}
&i\CM_{IV}^{ab,cd}=
 \CV_{\mu\nu\rho\rho'\lambda\lambda'}^{abdc}(k_{1},k_{2},-q_{2},-q_{1})~
e^{\mu}_{k_{1}}~e^{\nu}_{k_{2}}~e^{*\lambda\lambda'}_{q_{1}}~e^{*\rho\rho'}_{q_{2}}  ,
\eeqa
where $e^{\mu}_{k_{1}}$ is the wave function of the first vector boson
and $e^{\nu}_{k_{2}}$ - of the second.
The final tensor gauge bosons wave functions are $e^{*\lambda\lambda'}_{q_1}$
and $e^{*\rho\rho'}_{q_{2}}$.

The total amplitude is a sum of four terms:
\beqa\label{polarizedtransitionamplitude}
&i\CM  =  i\CM_{I}+ i\CM_{II}+i\CM_{III}+i\CM_{IV}.
\eeqa
Here we have been considering only the first nontrivial diagrams  Fig.\ref{fig5}-Fig.\ref{fig8},
but because the Lagrangian (\ref{Lagrangian}) contains also high-rank
tensor gauge fields of increasing order one should include their contribution
into the total amplitude as well. In the actual calculations we shall use the tensor
propagator (\ref{tensorpropagator}) with
\beqa\label{res}
\Pi_{\alpha\alpha'\beta\beta'} =
2(\eta_{\alpha\beta} \eta_{\alpha'\beta'}
+\eta_{\alpha \beta'}\eta_{\alpha' \beta}
-\eta_{\alpha \alpha'} \eta_{\beta\beta'})-
 {2\over 9}(\eta_{\alpha\beta}\eta_{\alpha'\beta'}
-\eta_{\alpha\beta'}\eta_{\alpha'\beta}),
\eeqa
which, as it appears, sums the diagrams with high-rank tensor fields in the
intermediate states\footnote{The details will be given elsewhere.}.
The total amplitude is gauge invariant, that is, if we take
longitudinal wave function for  gauge bosons, then
the transition amplitude vanishes  (see Appendix B for details).

Our intention is to calculate the physical matrix elements
in the helicity basis for initial vector and final tensor gauge bosons.
This calculation of polarized cross sections is very similar to the gluon-gluon
scattering in QCD \cite{feynman,Peskin:1995ev}.
The right- and left-handed  vector wave functions are:
\beqa \label{helicityvectors}
&e_{R}(k_{1})^{\mu}={1\over \sqrt{2}}(0,1,i,0)~~~~,~~~~
e_{L}(k_{1})^{\mu}={1\over \sqrt{2}}(0,1,-i,0)\nn
\\
&e_{R}(k_{2})^{\mu}=e_{L}(k_{1})^{\mu}~~~~,~~~~e_{L}(k_{2})^{\mu}=e_{R}(k_{1})^{\mu},
\eeqa
where $k_{1}^{\mu}=E(1,0,0,1),~~ k_{2}^{\mu}=E(1,0,0,-1)$
and the tensor gauge boson wave functions for circular polarizations along the
$\vec{q}_1 $ direction are
\beqa\label{tensorbosonhelicities}
&e_{R}^{\mu\alpha}(q_{1})={1\over 2}(0,\cos\theta,i,-\sin{\theta}) \bigotimes
(0,\cos\theta,i,-\sin{\theta}),~~~\nn\\
&e_{L}^{\mu\alpha}(q_{1})={1\over 2}(0,-\cos\theta,i,\sin{\theta}) \bigotimes
(0,-\cos\theta,i, \sin{\theta})
\eeqa
It is easy to check that the wave functions (\ref{tensorbosonhelicities}) are
orthonormal
$
e_{R}^{*\mu\alpha}(q_1)e_{L}(q_1)_{\alpha\nu }=0,~
$
$
e_{R}^{*\mu\alpha}(q_1)e_{R}(q_1)_{\mu\alpha}=1,~
$
$
e_{L}^{*\mu\alpha}(q_1)e_{L}(q_1)_{\mu\alpha}=1
$
and fulfil the equations
\beqa\label{polarizationidentities}
q^{\mu}_{1} e_{\mu \alpha}(q_1)=q^{\alpha}_{1} e_{\mu \alpha}(q_1)=0, ~~
q^{\mu}_{2} e_{\mu \alpha}(q_2)=q^{\alpha}_{2} e_{\mu \alpha}(q_2)=0,
\eeqa
\begin{figure}
\centerline{\hbox{\psfig{figure=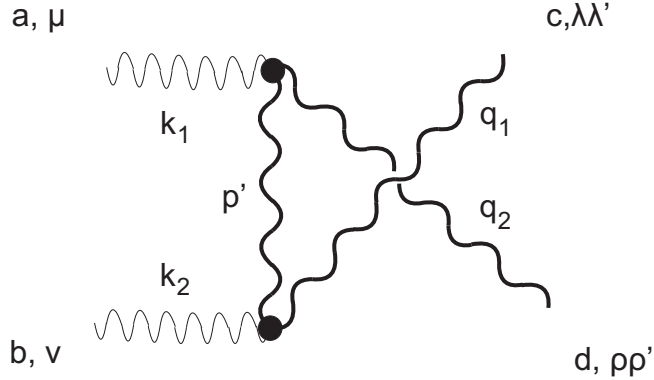,height=5cm,angle=0}}}
\caption[fig6]{The u-channel diagram for the process $V+V \rightarrow T+T$ .}
\label{fig6}
\end{figure}
The helicity states
for the second tensor gauge boson are
$
e_{R}^{\mu\nu}(q_{2})=e_{L}^{\mu\nu}(q_{1})~ ,~
e_{L}^{\mu\nu}(q_{2})=e_{R}^{\mu\nu}(q_{1}),
$
where $q_{1}^{\mu}=(E,E\sin\theta,0,E\cos\theta)$ and
$q_{2}^{\mu}=(E,-E\sin\theta,0,-E\cos\theta)$.

\section{\it Helicity Amplitudes}
Now we can calculate all sixteen matrix elements between states of definite
helicities.
The scattering amplitude (\ref{polarizedtransitionamplitude}) for any particular choice of helicities
contains four terms. By plugging explicit
expressions for propagators (\ref{vectorpropagator}), (\ref{tensorpropagator}), (\ref{res}),
vertices (\ref{VVV}),(\ref{VVVV}),(\ref{VTT}),(\ref{VVTT})  and helicity
wave functions (\ref{helicityvectors}),
(\ref{tensorbosonhelicities}) into the matrix elements
(\ref{polarizedtransitionamplitudeI}),(\ref{polarizedtransitionamplitudeII}),
(\ref{polarizedtransitionamplitudeIII}) and (\ref{polarizedtransitionamplitudeIV})
we can find their explicit form.
\begin{figure}
\centerline{\hbox{\psfig{figure=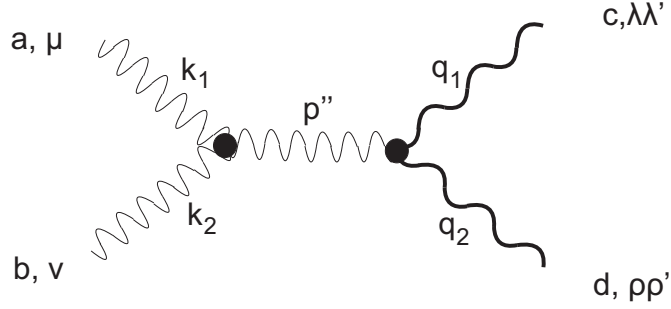,height=4cm,angle=0}}}
\caption[fig7]{The s-channel diagram for the process $V+V \rightarrow T+T$.}
\label{fig7}
\end{figure}
For the t-channel amplitude (\ref{polarizedtransitionamplitudeI}), corresponding
to the diagram Fig.\ref{fig5}, we shall get the following sequence
of sixteen polarization amplitudes
\beqa\label{vvtt1}
&i\CM_{I}(LL\rightarrow LL)=i\CM_{I}(RR\rightarrow RR)=-
g^{2}   {3i\over 4}f^{ace}f^{bde} (1+\cos\theta)(2-\cos{\theta})
\nn\\
&i\CM_{I}(LL\rightarrow LR)=i\CM_{I}(RR\rightarrow LR)=0\nn
\nn\\
&i\CM_{I}(LL\rightarrow RL)=i\CM_{I}(RR\rightarrow RL)=0\nn
\nn\\
&i\CM_{I}(LL\rightarrow RR)=i\CM_{I}(RR\rightarrow LL)=
-g^{2}   {i\over 8}f^{ace}f^{bde}(7-2\cos\theta+\cos{2\theta})
\nn\\
&i\CM_{I}(LR\rightarrow LL)=i\CM_{I}(LR\rightarrow RR)=
-g^{2}   {i \over 4} f^{ace}f^{bde}(1+\cos\theta)(3-2\cos\theta)
\nn\\
&i\CM_{I}(LR\rightarrow LR)=0,~~~~~
 i\CM_{I}(LR\rightarrow RL)=0
\nn\\
&i\CM_{I}(RL\rightarrow LL)=i\CM_{I}(RL\rightarrow RR)=
-g^{2}   {i \over 4} f^{ace}f^{bde}(1+\cos\theta)(3-2\cos\theta)
\nn\\
&i\CM_{I}(RL\rightarrow LR)=0,~~~~~
i\CM_{I}(RL\rightarrow RL)=0.
\eeqa
For the u-channel diagram Fig.\ref{fig6}  the  amplitude (\ref{polarizedtransitionamplitudeII})  gives
\beqa\label{vvtt2}
&i\CM_{II}(LL\rightarrow LL)=i\CM_{II}(RR\rightarrow RR)=
-g^{2}   {3i\over 4}f^{ace}f^{bde} (1-\cos\theta)(2+\cos{\theta})
\nn\\
&i\CM_{II}(LL\rightarrow LR)=i\CM_{II}(RR\rightarrow LR)=0
\nn\\
&i\CM_{II}(LL\rightarrow RL)=i\CM_{II}(RR\rightarrow RL)=0
\nn\\
&i\CM_{II}(LL\rightarrow RR)=i\CM_{II}(RR\rightarrow LL)=
-g^{2}   {i\over 8} f^{ade} f^{bce} (7+2\cos\theta-\cos{2\theta})
\nn\\
&i\CM_{II}(LR\rightarrow LL)=i\CM_{II}(LR\rightarrow RR)=
-g^{2}  {i\over 4}f^{ade}f^{bce}(1-\cos\theta) (3+2\cos\theta)
\nn\\
&i\CM_{II}(LR\rightarrow LR)=0,~~~~~
i\CM_{II}(LR\rightarrow RL)=0
\nn\\
&i\CM_{II}(RL\rightarrow LL)=i\CM_{II}(RL\rightarrow RR)=
-g^{2}   {i\over 4}f^{ade}f^{bce}(1-\cos\theta) (3+2\cos\theta)
\nn\\
&i\CM_{II}(RL\rightarrow LR)=0,~~~~~
i\CM_{II}(RL\rightarrow RL)=0.
\eeqa
For the s-channel diagram Fig.\ref{fig7} the polarization amplitudes
(\ref{polarizedtransitionamplitudeIII}) are
\beqa\label{vvtt3}
&i\CM_{III}(LL\rightarrow LL)=i\CM_{III}(RR\rightarrow RR)=
-g^{2}   {i \over 2} (f^{ace}f^{bde}-f^{ade}f^{bce}) \cos{\theta}
\nn\\
&i\CM_{III}(LL\rightarrow LR)=i\CM_{III}(RR\rightarrow LR)=0
\nn\\
&i\CM_{III}(LL\rightarrow RL)=i\CM_{III}(RR\rightarrow RL)=0
\nn\\
&i\CM_{III}(LL\rightarrow RR)=i\CM_{III}(RR\rightarrow LL)=
-g^{2}  {i \over 2} (f^{ace}f^{bde}-f^{ade}f^{bce}) \cos{\theta}
\nn\\
&i\CM_{III}(LR\rightarrow LL)=i\CM_{III}(LR\rightarrow RR)=0
\nn\\
&i\CM_{III}(LR\rightarrow LR)=0,~~~~~
i\CM_{III}(LR\rightarrow RL)=0
\nn\\
&i\CM_{III}(RL\rightarrow LL)=i\CM_{III}(RL\rightarrow RR)=0
\nn\\
&i\CM_{III}(RL\rightarrow LR)=0,~~~~~
i\CM_{III}(RL\rightarrow RL)=0.
\eeqa
\begin{figure}
\centerline{\hbox{\psfig{figure=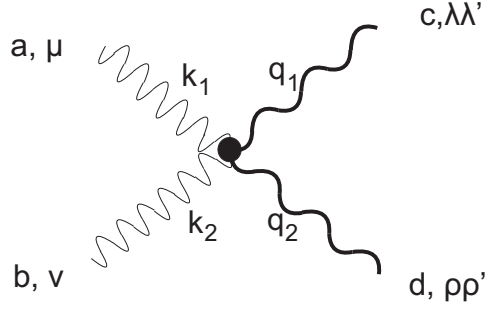,height=4cm,angle=0}}}
\caption[fig8]{The contact diagram for the process $V+V \rightarrow T+T$. }
\label{fig8}
\end{figure}
And finally for the contact diagram Fig.\ref{fig8} the polarization amplitudes
(\ref{polarizedtransitionamplitudeIV}) are
\beqa\label{vvtt4}
&i\CM_{IV}(LL\rightarrow LL)=i\CM_{IV}(RR\rightarrow RR)=
g^{2}   { i \over 4} (f^{ace}f^{bde}+f^{ade}f^{bce})\sin^{2}\theta
\nn\\
&i\CM_{IV}(LL\rightarrow LR)=i\CM_{IV}(RR\rightarrow LR)=0
\nn\\
&i\CM_{IV}(LL\rightarrow RL)=i\CM_{IV}(RR\rightarrow RL)=0
\nn\\
&i\CM_{IV}(LL\rightarrow RR)=i\CM_{IV}(RR\rightarrow LL)=
g^{2}   { i \over 4}(f^{ace}f^{bde}+f^{ade}f^{bce})\sin^{2}\theta
\nn\\
&i\CM_{IV}(LR\rightarrow LL)=i\CM_{IV}(LR\rightarrow RR)=
g^{2}   { i \over 4}(f^{ace}f^{bde}+f^{ade}f^{bce})\sin^{2}\theta
\nn\\
&i\CM_{IV}(LR\rightarrow LR)=0,~~~~~
i\CM_{IV}(LR\rightarrow RL)=0
\nn\\
&i\CM_{IV}(RL\rightarrow LL)=i\CM_{IV}(RL\rightarrow RR)=
g^{2}   { i \over 4}(f^{ace}f^{bde}+f^{ade}f^{bce})\sin^{2}\theta
\nn\\
&i\CM_{IV}(RL\rightarrow LR)=0,~~~~~~
i\CM_{IV}(RL\rightarrow RL)=0.
\eeqa
Thus only eight amplitudes out of sixteen are nonzero:
$
V_L V_L \rightarrow T_L T_L,~~V_R V_R \rightarrow T_R T_R,~~V_L V_L \rightarrow T_R T_R
,~~V_R V_R \rightarrow T_L T_L,~~V_L V_R \rightarrow T_L T_L,~~
V_L V_R \rightarrow T_R T_R,~~V_R V_L \rightarrow T_L T_L,~~V_R V_L \rightarrow T_R T_R.
$
We can get total helicity amplitudes (\ref{polarizedtransitionamplitude}) summing
corresponding terms from each diagram:
\vspace{0.5cm}
\beqa\label{polarizationLLLL}
i\CM_{LL\rightarrow LL}=-{i g^{2}\over 4}\biggl[&+&
f^{ace}f^{bde}
(4+5\cos\theta+\cos{2\theta} )+\nn \\
&+&f^{ade}f^{bce}
(4-5\cos\theta+\cos{2\theta} )\biggr],
\eeqa
\beqa\label{polarizationLLRR}
i\CM_{LL\rightarrow RR}=-{i g^{2}\over 4}\biggl[
f^{ace}f^{bde}
(3+\cos\theta)+
f^{ade}f^{bce}
(3-\cos\theta)\biggr],
\eeqa
\beqa\label{polarizationLRLL}
i\CM_{LR\rightarrow LL}=
-{i g^{2}\over 4}
\biggl[&+(1+\cos\theta)(2-\cos\theta)f^{ace}f^{bde}+ \nn\\
&+(1-\cos\theta)(2+\cos\theta)f^{ade}f^{bce}\biggr].
\eeqa
To compute the cross section, we must square matrix elements
(\ref{polarizationLLLL}), (\ref{polarizationLLRR}), (\ref{polarizationLRLL})
and then average
over the symmetries of the initial bosons and sum over the
symmetries of the final tensor gauge bosons. This gives
\beqa\label{squarepolarizationLLLL}
& {1\over d(G)^{2}}\sum_{col}|\CM_{LL\rightarrow LL}|^{2}=
 ~{g^{4}\over 32 }~ {C_{2}^{2}(G)\over d(G)}
 ( 124 - 23 \cos{2\theta}+3\cos{4\theta}  ),
\eeqa
\beqa\label{squarepolarizationLLRR}
&{1\over d(G)^{2}}\sum_{col}|\CM_{LL\rightarrow RR}|^{2}=
 ~ {g^{4} \over 32 } ~{C_{2}^{2}(G)\over d(G)}
(55+\cos{2\theta}),
\eeqa
\beqa\label{squarepolarizationLRLL}
&{1\over d(G)^{2}}\sum_{col}|\CM_{LR\rightarrow LL}|^{2}=
 ~{g^{4} \over 128 }~{C_{2}^{2}(G)\over d(G)}
(61-32\cos{2\theta}+3\cos{4\theta}),
\eeqa
where the invariant operator $C_2$ is defined by the equation
$ t^a t^a  = C_2~ 1 $.
We can calculate now the leading-order polarized cross sections for the
tensor gauge boson production $V+V \rightarrow T+T$.
\newline
{\it Helicity cross-sections.}
Plugging matrix elements (\ref{squarepolarizationLLLL}), (\ref{squarepolarizationLLRR}),
(\ref{squarepolarizationLRLL}), into our general
cross-section formula in the center-of-mass frame (\ref{crosssectionformula})
yields:
\beqa\label{crosssectionformulaLLLL}
d\sigma_{LL \rightarrow LL}  =
  {\alpha^2 \over  s   }   ~{  C^2_2(G) \over  128 d(G) }~
 ~ (124 -23 \cos{2\theta}+3\cos{4\theta})~ d\Omega ,
\eeqa
\beqa\label{crosssectionformulaLLRR}
d\sigma_{LL \rightarrow RR}  =
  {\alpha^2 \over  s   }   ~{  C^2_2(G) \over  128   d(G) }~
 ~ (55 + \cos{2\theta} )~ d\Omega ,
\eeqa
\beqa\label{crosssectionformulaLRLL}
d\sigma_{LR \rightarrow LL}  =
  {\alpha^2 \over  s   }   ~{  C^2_2(G) \over  512   d(G) }~
 ~  (61-32\cos{2\theta}+3\cos{4\theta})~ d\Omega ,
\eeqa
where
$
\alpha = {g^2 \over  4 \pi  }.
$
\newline
{\it Unpolarized cross section.}
Adding up all sixteen helicity amplitudes and dividing the result
by four in order to average over the initial boson spins we can get
unpolarized cross section. Thus summing over helicities
\beqa
& {1\over 4d(G)^{2}}\sum_{col,hel}|\CM|^{2}=
{1\over 4d(G)^{2}}\sum_{col} 2|\CM_{LL\rightarrow LL}|^{2}+
2|\CM_{LL\rightarrow RR}|^{2}+4|\CM_{LR\rightarrow RR}|^{2}=\nn \\
&=g_{2}^{2}
{g^{4}\over 128} {C_{2}^{2}(G) \over d(G)}
\biggl(~ 419 -76 \cos{2\theta}+9\cos{4\theta} ~\biggr)
\eeqa
for unpolarized cross section we shall get
\beqa\label{crosssectionunpolarized}
d\sigma   =
  {\alpha^2 \over  s   }   ~{  C^2_2(G) \over  d(G) }~
 ~  { 419-76\cos{2\theta}+9\cos{4\theta} \over  512  }~ d\Omega ,
\eeqa
where for the $SU(N)$ group we have ${ C^{2}_2(G) \over    d(G) }~
 ~=~{   N^2   \over  (N^2 -1)  }$.
The production cross section of tensor
gauge bosons (\ref{crosssectionunpolarized}) has characteristic
behavior with its maximum at $\theta=\pi/2$ and decrease for small angles.

This cross section should be compared with the analogous  cross section
in QCD. Indeed, let us compare this result with the gluon-gluon scattering \cite{Peskin:1995ev}.
The $V+V \rightarrow V+V$  cross section  is
\beqa\label{VVVVscattering}
d \sigma   = {\alpha^{2}\over s} ~{C_{2}^{2}(G) \over d(G)}~
{(4-sin^{2}\theta)^{3} \over 32 \sin^{4}\theta}~ d\Omega.
\eeqa
This cross section increases at small angles $\theta \sim 0, \pi$ and therefore the scattering
is mostly going into forward and backward directions and has its minimum in transverse direction
at $\theta=\pi/2$. The production cross section of spin-two
gauge bosons (\ref{crosssectionunpolarized}) shows dramatically
different behavior  with its maximum in the transverse direction
at $\theta=\pi/2$ and decrease in forward and backward directions.
One can only speculate that at high enough energies, may be at LHC energies,
we may observe the standard
spin-one gauge boson together with its new partner, spin-two gauge boson.

\section{\it Appendix A}

Here we shall review the well known result for the three-level gluon scattering
$V+V \rightarrow V+V$ \cite{Peskin:1995ev} in order to compare it with the tensor scattering
considered above: $V+V \rightarrow T+T$.
Because of the parity and crossing symmetry from 16 possible amplitudes only 5
are different. We have the following equalities:
\beqa
&i\CM(LL\rightarrow LL)=i\CM(RR\rightarrow RR),~~~
i\CM(LL\rightarrow RR)=i\CM(RR\rightarrow LL),\nn\\
&i\CM(LL\rightarrow LR)=i\CM(RR\rightarrow LR)=i\CM(LR\rightarrow LL)=i\CM(LR\rightarrow RR)=\nn \\
=&i\CM(LL\rightarrow RL)=i\CM(RR\rightarrow RL)=
i\CM(RL\rightarrow LL)=i\CM(RL\rightarrow RR),\nn\\
&i\CM(LR\rightarrow LR)=i\CM(RL\rightarrow RL),~~~
i\CM(LR\rightarrow RL)=i\CM(RL\rightarrow LR).\nn
\eeqa
Four Feynman diagram contribute into this scattering. The contribution
of the t-channel diagram can be expressed in the form:
\beqa\label{vvvv1}
&i\CM_{I}(LL\rightarrow LL)={ig^2\over 8}f^{ace}f^{bde}
(39-24\cos\theta+\cos{2\theta})\cot^{2}{\theta\over 2}
\nn\\
&i\CM_{I}(LL\rightarrow LR)={ig^2\over 4}f^{ace}f^{bde}\sin^{2}\theta
\nn\\
&i\CM_{I}(LL\rightarrow RR)={ig^2\over 2}f^{ace}f^{bde}
\biggl({3+\cos\theta\over \sin^{2}{\theta\over 2}}\biggr)\sin^{4}{\theta\over 2}
\nn\\
&i\CM_{I}(LR\rightarrow LR)={ig^2\over 2}f^{ace}f^{bde} \biggl({3+\cos\theta\over
\sin^{2}{\theta\over 2}}\biggr)\cos^{4}{\theta\over 2}
\nn\\
&i\CM_{I}(LR\rightarrow RL)={ig^2\over 2}f^{ace}f^{bde} \biggl({3+\cos\theta\over
\sin^{2}{\theta\over 2}}\biggr)\sin^{4}{\theta\over 2},
\eeqa
of the u-channel diagram as
\beqa\label{vvvv2}
&i\CM_{II}(LL\rightarrow LL)={ig^2\over 8}f^{ade}f^{bce}(39+24\cos\theta+\cos{2\theta})\tan^{2}{\theta\over 2}
\nn\\
&i\CM_{II}(LL\rightarrow LR)={ig^2\over 4}f^{ade}f^{bce}\sin^{2}\theta
\nn\\
&i\CM_{II}(LL\rightarrow RR)={ig^2\over 2}f^{ade}f^{bce} \biggl({3-\cos\theta\over \cos^{2}{\theta\over 2}}\biggr)\cos^{4}{\theta\over 2}
\nn\\
&i\CM_{II}(LR\rightarrow LR)={ig^2\over 2}f^{ade}f^{bce} \biggl({3-\cos\theta\over \cos^{2}{\theta\over 2}}\biggr)\cos^{4}{\theta\over 2}
\nn\\
&i\CM_{II}(LR\rightarrow RL)={ig^2\over 2}f^{ade}f^{bce} \biggl({3-\cos\theta\over \cos^{2}{\theta\over 2}}\biggr)\sin^{4}{\theta\over 2},
\eeqa
of the s-channel  diagram as
\beqa\label{vvvv3}
&i\CM_{III}(LL\rightarrow LL)=-ig^2 f^{abe}f^{cde}\cos\theta
\nn\\
&i\CM_{III}(LL\rightarrow LR)=0
\nn\\
&i\CM_{III}(LL\rightarrow RR)=-ig^2 f^{abe}f^{cde}\cos\theta
\nn\\
&i\CM_{III}(LR\rightarrow LR)=0
\nn\\
&i\CM_{III}(LR\rightarrow RL)=0
\eeqa
and of the contact diagram as
\beqa\label{vvvv4}
&i\CM_{IV}(LL\rightarrow LL)=-ig^{2}\biggl[f^{abe}f^{cde}
\cos\theta+f^{ace}f^{bde}(1-\sin^{4}{\theta\over 2})+
f^{ade}f^{bce}(1-\cos^{4}{\theta\over 2})\biggr]
\nn\\
&i\CM_{IV}(LL\rightarrow LR)=-{ig^{2}\over 4}
(f^{ace}f^{bde}+f^{ade}f^{bce})\sin^{2}\theta
\nn\\
&i\CM_{IV}(LL\rightarrow RR)=-ig^{2}\biggl[f^{abe}f^{cde}\cos\theta+f^{ace}f^{bde}(1-\cos^{4}{\theta\over 2})+
f^{ade}f^{bce}(1-\sin^{4}{\theta\over 2})\biggr]
\nn\\
&i\CM_{IV}(LR\rightarrow LR)=ig^{2}(f^{ace}f^{bde}+f^{ade}f^{bce})\cos^{2}\theta
\nn\\
&i\CM_{IV}(LR\rightarrow RL)=ig^{2}(f^{ace}f^{bde}+f^{ade}f^{bce})\sin^{2}\theta.
\eeqa
So for the total amplitudes we have
\beqa
i\CM_{LL\rightarrow LL}=4ig^{2}\biggl[\biggl({1\over 1-\cos\theta}\biggr)
f^{ace}f^{bde}+\biggl({1\over 1+\cos\theta}\biggr)f^{ade}f^{bce}\biggl]
\eeqa
\beqa
i\CM_{LR\rightarrow LR}=2ig^{2}\cos^{2}{\theta\over 2}\cot{\theta\over 2}
\biggl(\cot{\theta\over 2}f^{ace}f^{bde}+\tan{\theta\over 2}f^{ade}f^{bce}\biggr)
\eeqa
\beqa
i\CM_{LR\rightarrow RL}=ig^{2}(1-\cos\theta)^{2}
\biggl[\biggl({1\over 1-\cos\theta}\biggr)f^{ace}f^{bde}+
\biggl({1\over 1+\cos\theta}\biggr)f^{ade}f^{bce}\biggl]
\eeqa
together with
$
i\CM_{LL\rightarrow LR}=0
$,
$
i\CM_{LL\rightarrow RR}=0
$. Thus only three of 16 helicity amplitudes are nonzero. Squaring the matrix
elements one can get
\beqa\label{sqvvvv1}
|\CM_{LL\rightarrow LL}|^{2}={16g^{4}\over d(G)}C_{2}^{2}(G)
\biggl({3+\cos^{2}\theta\over \sin^{4}\theta}\biggr),
\eeqa
\beqa\label{sqvvvv2}
|\CM_{LR\rightarrow LR}|^{2}={g^{4}\over d(G)}C_{2}^{2}(G)
\biggl({3+\cos^{2}\theta\over \sin^{4}\theta}\biggr)(1+\cos\theta)^{4},
\eeqa
\beqa\label{sqvvvv3}
|\CM_{LR\rightarrow RL}|^{2}={g^{4}\over d(G)}C_{2}^{2}(G)
\biggl({3+\cos^{2}\theta\over \sin^{4}\theta}\biggr)(1-\cos\theta)^{4}.
\eeqa
Using these formulas and (\ref{crosssectionformula}) one can easily
get the cross section (\ref{VVVVscattering}). It is also
instructive to compare the above helicity amplitudes (\ref{vvvv1})-(\ref{vvvv4})
with the corresponding helicity amplitudes for the tensor gauge bosons
(\ref{vvtt1})-(\ref{vvtt4}). The characteristic feature of the squared amplitudes (\ref{sqvvvv1}) -
(\ref{sqvvvv3}) is that they increase at small angles $\theta \sim 0, \pi$ and therefore the scattering
is mostly going into forward and backward directions. In contrast to that behavior tensor gauge
boson amplitudes (\ref{squarepolarizationLLLL})-(\ref{squarepolarizationLRLL}) are  decreasing
at $\theta \sim 0, \pi$  and increasing in the transverse direction $\theta \sim  \pi/2$.

\section{\it Appendix B}

To check on mass-shell gauge invariance of the amplitude (\ref{polarizedtransitionamplitude})
let one of the tensor gauge boson wave function be longitudinal:
$$e^{\rho\rho'}_{q_{2}} =q_{2}^{\rho}\xi^{\rho'}+q_{2}^{\rho'}\xi^{\rho}.$$
On mass-shell gauge transformations  should
fulfill the following conditions:  $q^2_{2}=0,~q_2 \cdot e_{q_2} =0,~tre_{q_2}=0.$
These equations are satisfied if
$ q_{2} \cdot \xi =0$ and therefore
\[
\xi_{0}=\xi_{1}\sin\theta+\xi_{3}\cos\theta,
\]
where  $q_{2}^{\mu} = E (1,-sin\theta,0,-\cos\theta)$. To see explicitly how
the cancelation between diagrams takes place let us take the rest of the
vector bosons left polarized and one of the tensor bosons
right polarized. In that case we shall get the following amplitudes:
\beqa
i\CM^{long}_{I} =ig^{2} f^{ace}f^{bde}
 E ~\cos^{3}{\theta\over 2}
\biggl[(\xi_{0}-\xi_{3})\cos{\theta\over 2}-(\xi_{1}+i\xi_{2})
\sin{\theta\over 2}\biggr]
 \equiv  f^{ace}f^{bde} M_{1}, \nn
\eeqa
\beqa
i\CM^{long}_{II} =ig^{2} f^{ade}f^{bce}
 E~ \sin^{3}{\theta\over 2}
\biggl[(\xi_{0}+\xi_{3})\sin{\theta\over 2}-
(\xi_{1}-i\xi_{2})\cos{\theta\over 2}\biggr]
 \equiv f^{ade}f^{bce}  ~M_{2},  \nn
\eeqa
\beqa
i\CM^{long}_{III} &=&-{ig^{2} \over 4}(f^{ace}f^{bde}-f^{ade}f^{bce})
E~\sin\theta(\xi_{1}\cos\theta-i\xi_{2}-\xi_{3}\sin\theta)\nn \\
& \equiv & (f^{ace}f^{bde}-f^{ade}f^{bce}) ~M_{3}, \nn
\eeqa
\beqa
i\CM^{long}_{IV} &=&{ig^{2} \over 4}(f^{ace}f^{bde}+f^{ade}f^{bce})
E~\sin\theta\cos\theta(i\xi_{2}-\xi_{1}\cos\theta+\xi_{3}\sin\theta)\nn \\
&\equiv&(f^{ace}f^{bde}+f^{ade}f^{bce})~E~M_{4},\nn
\eeqa
where
\beqa\label{longitudinalparts}
M_{1} &=& ig^{2} E~\cos^{3}{\theta\over 2}
\biggl[(\xi_{0}-\xi_{3})\cos{\theta\over 2}-(\xi_{1}+i\xi_{2})\sin{\theta\over 2}\biggr],\nn\\
M_{2}&=&ig^{2}  E~\sin^{3}{\theta\over 2}
\biggl[(\xi_{0}+\xi_{3})\sin{\theta\over 2}-(\xi_{1}-i\xi_{2})\cos{\theta\over 2}\biggr], \nn\\
M_{3}&=&-{ig^{2} \over 4}E~\sin\theta(\xi_{1}\cos\theta-i\xi_{2}-\xi_{3}\sin\theta),\nn\\
M_{4}&=&{ig^{2} \over 4}E~\sin\theta\cos\theta(i\xi_{2}-\xi_{1}\cos\theta+\xi_{3}\sin\theta).
\eeqa
For the total amplitude we shall get
\beqa
i\CM^{long}&=& f^{ace}f^{bde}
\biggl[ M_{1}+M_{3}+M_{4}    \biggr] +
f^{ade}f^{bce}\biggl[ M_{2}-M_{3}+M_{4}   \biggr],
\eeqa
and it nullifies if
\beqa
& M_{1}+M_{3}+M_{4}=0,~~~~~~ M_{2}-M_{3}+M_{4}=0
\eeqa
or, equivalently:
$
M_{1}+M_{2}+2M_{4}=0, ~~
 M_{1}-M_{2}+2M_{3}=0.
$
Using explicit expressions (\ref{longitudinalparts}) one can check that these equations
are identically satisfied for arbitrary functions $\xi_{1},\xi_{2},\xi_{3}$.

The work of (G.S.) was partially supported by ENRAGE (European Network on Random
Geometry), Marie Curie Research Training Network, contract MRTN-CT-2004-
005616.

\vfill
\end{document}